\documentclass[aps,prd,twocolumn,nofootinbib,showpacs,superscriptaddress,groupedaddress,amsmath]{revtex4-1}  
\usepackage{graphicx}  
\usepackage{dcolumn}   
\usepackage{bm}        
\usepackage{amssymb}
\usepackage{subfig}
\usepackage{url}
\usepackage{filecontents}
\usepackage{natbib}
\usepackage{aas_macros}

\usepackage{xspace}
\usepackage{url}
\usepackage{wasysym}
\def\fermi{{\it Fermi}-LAT\xspace}

\def\gray{$\gamma$-ray\xspace}

\begin{document}
\title{Radial distribution of  the diffuse gamma-ray emissivity in the galactic disk}
\author{Ruizhi Yang}\email{ryang@mpi-hd.mpg.de}
\affiliation{%
\ Max-Planck-Institut f{\"u}r Kernphysik, P.O. Box 103980, 69029 Heidelberg, Germany\\
}
\author{Felix Aharonian}\email{felix.aharonian@mpi-hd.mpg.de}
\affiliation{%
\ Max-Planck-Institut f{\"u}r Kernphysik, P.O. Box 103980, 69029 Heidelberg, Germany\\
\ Dublin Institute for Advanced Studies, 31 Fitzwilliam Place, Dublin 2, Ireland.
}
\author{Carmelo Evoli}\email{carmelo.evoli@gssi.infn.it}
\affiliation{%
\ Gran Sasso Science Institute, 7 viale Francesco Crispi, 67100 LÕAquila (AQ), Italy
}

\date{Received:  / Accepted: } 

\begin{abstract}
The \fermi  data accumulated over 7 years of  \gray observations,  together with 
the high resolution  gas (CO \& HI)  and  the  dust opacity maps, are used  to study the  
emissivity  of $\gamma$-rays induced  by interactions of cosmic rays  (CRs)  with the interstellar medium. 
Based on the  dust opacity templates,   the  \gray emissivity  was  measured for  36 segments  of the Galactic plane. 
Furthermore,   the \gray emissivity   was  evaluated  in six Galactocentric rings. Both the absolute emissivity  and the energy spectra of $\gamma$-rays  derived in the interval 0.2-100~GeV 
show  significant  variations along the galactic plane.   
The density of CRs,  derived under the assumption that  $\gamma$-rays  are  predominately produced in  
CR interactions with the interstellar gas,   is characterised by a strong radial dependence.  In the inner Galaxy  
the CR density substantially exceeds  the  density in the outer parts of the Galaxy: by  a   factor of  a few  at 10 GeV, 
and by more than  an order of magnitude at 1~TeV.   Remarkably, the  energy distribution  of CRs appears to be 
substantially harder  than the energy spectrum  obtained from direct measurements of local CRs. 
At the same time,  the flux and the energy spectrum of multi-GeV protons derived from $\gamma$-ray data 
in  the outskirts of the Galaxy is  quite close to the measurements of  local CRs.

\end{abstract}
\maketitle\section{Introduction}
There is a general consensus within the cosmic ray (CR) physics community 
that  up to  the so-called  \emph{knee}, a distinct spectral feature  around $10^{15}$~eV, CRs are 
produced in  galactic sources, presumably  in  supernova remnants \citep[for a review, see e.g.][]{drury12,blasi13,cristofari13}
The direct studies  of  CRs, which include  measurements  of  the energy spectrum, the mass composition and  
arrival directions of  particles,  provide an important  but yet not decisive  information about  the 
production sites of CRs  and their  propagation  in  the Galaxy.   
The  diffusion of CRs in  turbulent  magnetic  fields erases, to a large extent,  the information on 
the distribution of CR accelerators.  The energy dependent-diffusion of 
CRs also significantly modifies the initial (acceleration) spectra of CRs. 
In this regard,  $\gamma$-rays,  the secondary   products of interactions of CRs,  tell us more
information about the  origin of CRs.  Namely, while the discrete \gray  sources elucidate  the 
locations  of  individual CR accelerators, the diffuse \gray  component  contains an 
information about the global distribution of  CR factories, as well as about 
the CR propagation  throughout  the Galaxy. 

Historically, since the  first \gray  measurements,  the study of the diffuse gamma-radiation of the Galactic Disk 
has been  among  the major objectives of  gamma-ray astronomy.   Unfortunately, 
this  component of radiation  is contaminated by contributions of  weak but numerous unresolved \gray sources. 
Therefore its extraction is a  quite difficult task.  This can explain  the relatively slow progress  
in  this area. The 
improvement of the  flux sensitivity by  \fermi, compared to the previous 
gamma-ray missions,  by an order of magnitude, and, perhaps more importantly, 
the extension of the dynamical range of 
measurements over more than three decades, from  100 MeV to $\geq 100$~GeV,  promise a breakthrough 
regarding both the removal of the background caused by individual  sources and separation of 
the leptonic (bremsstrahlung and inverse Compton) and hadronic ($pp$ and $pA$) radiation components. 

Over the  last several years,  a significant progress has been achieved  also towards the 
knowledge of the gas distribution in our Galaxy. This is  a key issue for extraction  of the 
diffuse galactic \gray emission and for derivation of the spectral and spatial distributions of  
parent CR particles.    In this regard, one should mention 
the recent reports of the  Planck Collaboration on  the dust opacity maps which are 
complementary to the CO and 21 cm  measurements of the  molecular and atomic  
hydrogen distributions, respectively.  

In this paper,  we report   the results of our study of the diffuse \gray  emission of the Galactic Disk 
based  on the \fermi  observations,  and briefly discuss the implications  of these results,
in particular  in the context of spatial and energy distributions of CRs in the Galactic Disk.

\section{The gas tracers}
 The traditional tracers of the hydrogen in the atomic and   
molecular forms are the  21 cm HI  and 2.6 mm CO lines, respectively.  In this paper we use the data from CO galactic survey of  \citet{dame01} with the 
CfA 1.2m millimetre-wave Telescope,  and the Leiden/Argentine/Bonn (LAB) Survey on HI gas. 
For the CO  data, we use the standard assumption of a linear relationship between the velocity-integrated 
CO intensity, $W_{\rm CO}$, and the column density of molecular hydrogen, N(H$_{2}$). 
The conversion factor $X_{\rm CO}$ may be different in different parts of the galaxy,  therefore 
we leave this quantity as a free parameter in the likelihood fitting below. 

For the HI data  we use the equation 
\begin{equation}
N_{HI}(v,T_s)=-log \left(1-\frac{T_B}{T_s-T_{bg}}\right)T_sC_i\Delta v \ ,
\end{equation}
where $T_{bg}\approx2.66$~K is the brightness temperature of the cosmic microwave background radiation at 21cm, and
$C_i = 1.83 \times 10^{18} \rm  cm^{2}$.  In  the case  when $T_B > T_s-5~\rm K$, 
we truncate $T_B$ to $T_s-5~\rm K$; 
$T_s$ is chosen to be 150 K. The systematic uncertainties due to the different spin temperatures 
are discussed in ref.\citet{fermi_diffuse}. The effect, however,  is  quite small in most regions of the sky.

For  different  reasons,  the neural gas  cannot  be always  traced by  CO and HI observations. 
In such  cases (e.g. in optically thick clouds),  the  infrared emission from cold interstellar 
dust provides an alternative and  independent  measurements  of the gas column density.  
To find  it, we need a relation between the dust opacity and the column density.  
According to Equation~(4) of ref.\citet{planck}, 
\begin{equation}\label{eq:dust}
\tau_M(\lambda) = \left(\frac{\tau_D(\lambda)}{N_H}\right)^{dust}[N_{H{\rm I}}+2X_{CO}W_{CO}],
 \end{equation}
where $\tau_M$ is the dust opacity as a function of the wavelength  $\lambda$,  $(\tau_D/N_H)^{dust}$ is the reference dust emissivity measured in low-$N_H$ regions, $W_{CO}$ is the integrated brightness temperature of the CO emission, and $X_{CO}=N_{H_{2}}/W_{CO}$ is the  $H_2/CO$ conversion factor.
The substitution of  the latter into Equation~(\ref{eq:dust})  gives 
\begin{equation}
N_H = N_{H{\rm I}} +2 N_{H_2} =  \tau_m(\lambda)\left[\left(\frac{\tau_D(\lambda)}{N_H}\right)^{dust}\right]^{-1}. 
\end{equation}
Here  for the dust emissivity at $353~\rm GHz$,   we use  $(\tau_D/N_H)^{dust}_{353{\rm~GHz}}=1.18\pm0.17\times10^{-26}$~cm$^2$  taken from Table~3 of  ref.\citet{planck}. 

General,  the dust opacity is considered  as a robust and reliable estimate of the gas column density. 
On the other hand, the dust opacity maps do not contain an  information on the distance. 
In order to investigate the Galactocentric radial distribution of CRs we need to divide 
the gas distribution into different rings around the  GC.  For this purpose  we use the following  relation 
\begin{equation}
V_{LSR}=R_{\odot}( \frac{V(R)}{R}-\frac{V_{\odot}}{R_{\odot}})sin(l)cos(b),
\label{eq:rot}
\end{equation}
where $R$ is the galactocentric distance, $V(R)$ is the Galactic rotational curve,  and $l$ and $b$  are
the galactic coordinates in the line  of sight (LOS). We adopt the rotational curve parametrised in ref.\citet{clemens85}; $V_{\odot}$ and $R_{\odot}$ are fixed to 220 km/s and 8.5 kpc, respectively. By applying this to both CO and HI data, we can transform the velocity information into galactocentric distance of the gas. Below the CO and HI data are binned into six distance intervals:  [1,2], [2,4],[4,6],[6,8],[8,12 ] and [12,25] kpc. Due to the limited sky coverage  by  the CO surveys,
we limit our study by the regions with $|b|<5^{\circ}$. It should be noted that  Eq.\ref{eq:rot}  
allows  emission  from  the forbidden velocity zones in the CO and HI data. 
The reason could be  the non-circular motion of gas. 
This component contains only small fraction of total gas, therefore for simplicity we assigned it into the local rings. 
In the  GC and anti-center directions, the velocity resolution  does not allow us to determine  the distance. 
Therefore in the gamma ray likelihood  fittings we mask out  the 10 degree regions  around  the GC and anti-center. 

As mentioned above, in some cases the HI and CO measurements cannot  trace the neutral gas. For the missing "dark" gas,
we  use the dust data to derive the residual templates. We fit the dust opacity maps as a linear combination of HI and CO maps, and, in this way, find   the residual map as the "dark" gas template. Then we iterate this fit by including the "dark" gas template until  a convergence is achieved . To account for the $X_{CO}$ variation in different regions of the Galaxy, 
we perform the fitting in different segments  of the galactic plane, each of  $10^{\circ} \times 10^{\circ}$ angular size. 
This method is similar to the derivation of  the $E(B-V)_{res}$ templates used by the Fermi LAT collaboration \citet{fermi_diffuse},  
but  instead of the extinction maps we use the dust opacity maps. The detailed study of the "dark" gas distribution is beyond the scope of this paper.  But, in any case, the gamma-ray fitting below shows  that in our ROI the "dark"  gas has only 
minor impact on the final results.

\section{gamma ray  data}

In this study we  included  the observations of  \fermi  accumulated over a period of approximately 7 years (MET 239557417 -- MET 451533077), and selected all events with energies  above 100~MeV. 
For the data reduction, we use the standard LAT analysis software package \emph{v10r0p5}\footnote{\url{http://fermi.gsfc.nasa.gov/ssc}}. 
%
%
%
In order to reduce the effect of the background caused by  the Earth albedo, we excluded from the analysis the time intervals when the Earth was in the field-of-view (more specifically,  when the centre of the field-of-view was more than $52^ \circ$ above the zenith), as well as the time intervals when the parts of the ROI were observed at zenith angles $> 90^ \circ$. 
The spectral analysis was performed based on the P8R2\_v6 version of the post-launch instrument response functions (IRFs). Both the front and back converted photons have been selected.
Since the galactic diffuse model provided by the Fermi collaboration\footnote{gll\_iem\_v06.fit, available at \url{http://fermi.gsfc.nasa.gov/ssc/data/access/lat/BackgroundModels.html}} already contains the  emission component 
from the CR-gas interactions,  we do not include it in the analysis.  In the Galactic  plane,  in the energy interval of interest between 100~MeV and 100~GeV,  the diffuse \gray emission is dominated by  the channel of decay of neutral pions produced   
at interactions of CR protons and nuclei  with the interstellar gas, with a significantly smaller contribution from the Inverse Compton (IC) scattering of relativistic electrons \citep[see e.g.][]{aa00}.  The emissivity of the pion decay gamma rays is proportional to both the CR and gas densities. Thus we use the dust opacity map and gas maps  based  on the CO and HI observations  as the  ``pion-decay template". The IC templates has been calculated by using the 
Galprop code\footnote{\url{http://galprop.stanford.edu/webrun/}} \citep{galprop}, based on the information 
regarding the CR electrons and the interstellar radiation field (ISRF). Furthermore, we included an  
isotropic template to take into account the foregrounds like 
the contamination  caused  directly by  CRs,  and  by the  isotropic extragalactic \gray  background. 
Finally, we used the 3FGL catalogue  \citep{3fgl}  of the \fermi collaboration, to exclude contributions from the 
the resolved point-like \gray sources.

For  evaluation of  CR density  in different regions of the Galactic Disk, the standard \fermi tools {\it gtlike} are
not suitable for this task. To perform the analysis,  we first divide the entire  0.1-200 GeV region into 
10 equal in logarithmic scale  energy bins, and generate the count maps and exposure maps in each energy bin by using the standard routine  {\it gtbin} and {\it gtexpcube2}. Then we generate the source map by convolving the point sources and the diffuse emission templates with the point spread function (PSF) in each energy bin.  At low energies, to account  for the larger PSF,
we produce  the source map in a larger area compared with the ROI. For example,  in the analysis below with dust templates,  where each ROI is a $10^{\circ}\times10^{\circ}$ square, 
we generate the source map for diffuse templates with a size of $30^{\circ}\times30^{\circ}$.  The likelihood function has the same form as the one used in {\it gtlike},  $log(L)=\sum_{k_i}log(\mu_i)-\mu_i-log(k_i!)$, where $k$ is the number of photons in the $i$-th bin in the  counts map, and $\mu_i$ is the predicted number of photons within a particular linear combination of the templates. We use {\it iminuit} \footnote{\url{http://pypi.python.org/pypi/iminuit}} to minimise  the -log(L) function and derive the best fit parameter and their error distributions. 

\subsection {The analysis with dust templates}
\begin{figure*}
\includegraphics[width=0.5\linewidth]{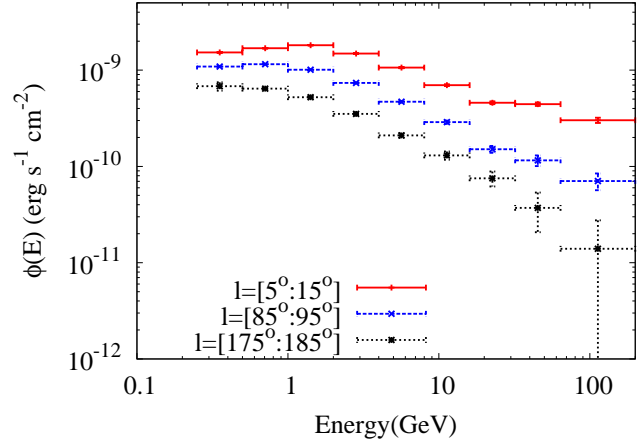}
\caption{The spectral energy distribution (SED) of the galactic diffuse \gray emission associated with the 
dust opacity in three different directions.}
\label{fig:dustsed}
\end{figure*}

\begin{figure*}
\includegraphics[width=0.5\linewidth]{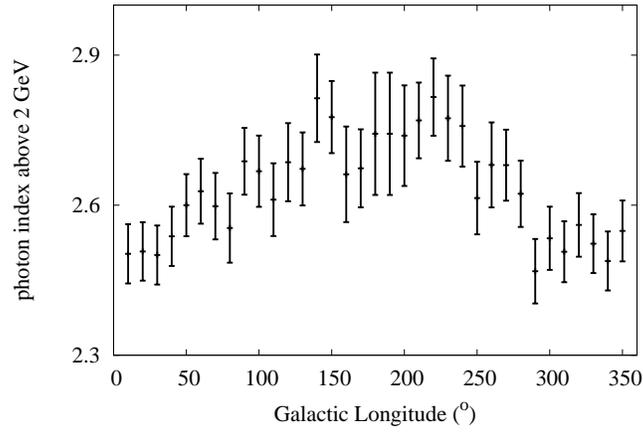}
\caption{The distribution of  the   power-law photon index of the galactic diffuse \gray emission 
associated with the dust opacity over the galactic longitudes integrated for the interval   $|b|<5^{\circ}$.}
\label{fig:dustindex}
\end{figure*}

We divided the plane into thirty six   $10^{\circ} \times10^{\circ} $ segment. In each patch, we fit the  \gray emission 
using  the catalog point sources, the dust opacity template, the IC template and the isotropic template. To deal with the point sources, 
we fit all  \gray point sources  with TS value larger than 25 in the radius of interest, 
and use the catalog spectral information for  sources with smaller TS.
The results are summarised  in Fig.\ref{fig:dustsed} and Fig.\ref{fig:dustindex}.  

In Fig.\ref{fig:dustsed}  are shown the  spectral energy distributions  (SED),  $E^2  {\rm d} N/{\rm d}E$,  characterising 
the energy spectra of  $\gamma$-rays  from  three different regions of the Galactic Disk. 
They are similar in a sense that above 2 GeV all three  can be approximated by a power-law distribution,
${\rm d} N/{\rm d}E \propto E^{-\Gamma}$,  and below 2 GeV they are flat. At the same time, these spectra differ 
from each other, namely  at high energies they are characterised by different  values of the 
photon index $\Gamma$.  The strong variation
of the spectral index  in the  galactic plane is demonstrated in Fig.\ref{fig:dustindex}.  Note that while  at energies 
above 2 GeV the photon index $\Gamma$ mimics the  power-law index  of the spectrum of parent protons, 
the hardening of the spectrum below 2 GeV 
is explained by the specifics of the kinematics of  decay of  secondary $\pi^0$ mesons from  $pp$ interactions, and, 
partly,   by contributions of  leptonic components of  $\gamma$-rays from   bremsstrahlung and inverse 
Compton scattering of relativistic electrons \citep[see e.g.][]{aa00}.  
The main conclusion from  Fig.\ref{fig:dustsed} and  Fig.\ref{fig:dustindex}
is that the photon  indices  of the diffuse  $\gamma$-ray emission associated  with the pion-decay $\gamma$-rays, 
increases  significantly from 2.4 in the GC direction to 2.8 in the anti-GC direction. 
This  can be interpreted as  a  tendency  of hardening of the spectrum of CR protons and nuclei 
in the inner Galaxy.

\begin{figure*}
\includegraphics[width=0.5\linewidth]{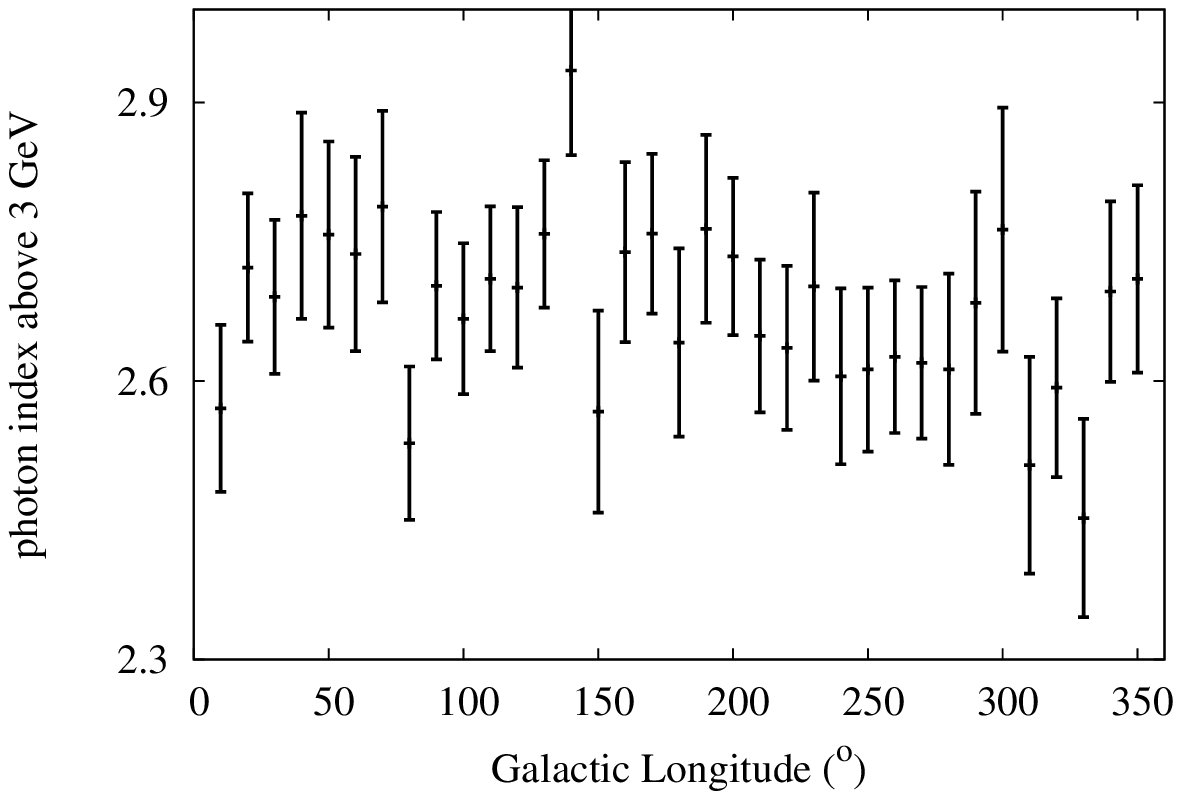}
\caption{The same as in Fig.\ref{fig:dustindex} but for the latitude interval  $5^{\circ}<|b|<10^{\circ}$.}
\label{fig:index10}
\end{figure*}

\begin{figure*}
\includegraphics[width=0.5\linewidth]{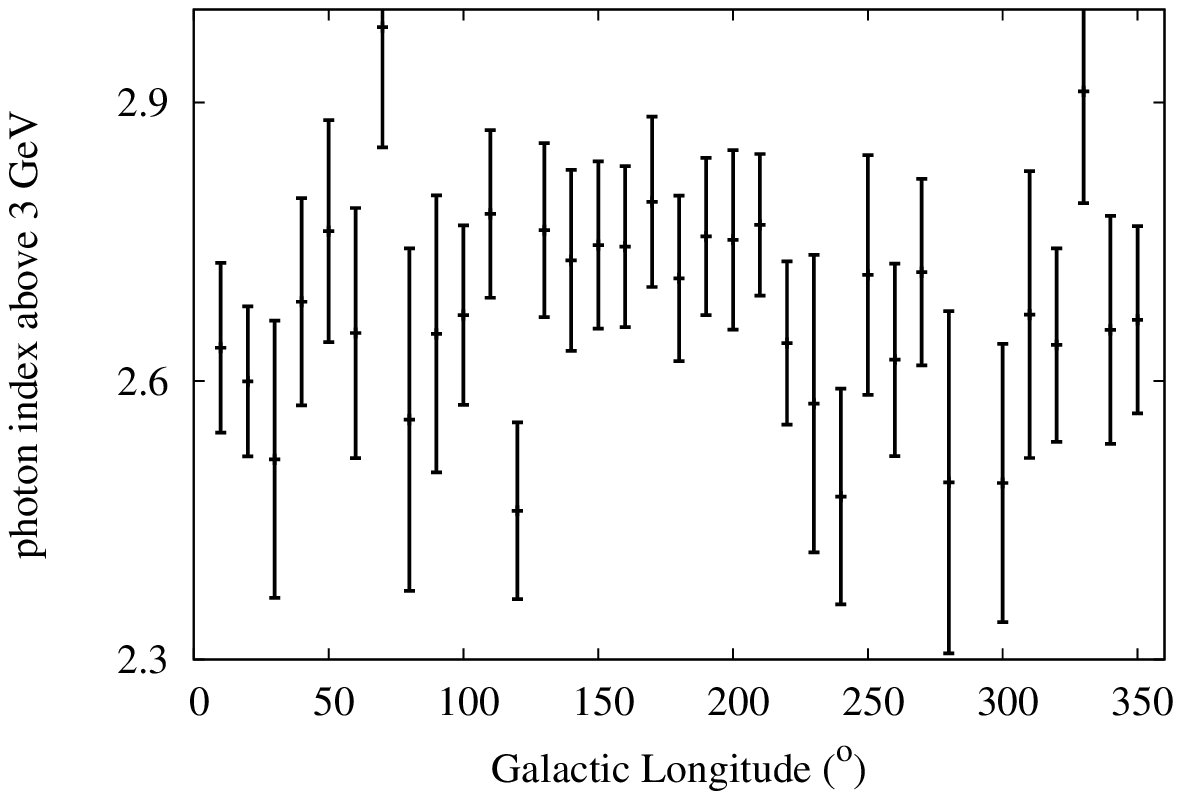}
\caption{The same as in Fig.\ref{fig:dustindex}  but for the latitude interval 
$10^{\circ}<|b|<15^{\circ}$.}
\label{fig:index15}
\end{figure*}

For a cross check,  we also apply {\it gtlike} in the analysis with dust templates. 
The results appear to be  consistent with that from the likelihood analysis. 
We also check whether this effect appears also at higher latitudes. To do this, we performed  the same analysis in the latitude intervals  $5^{\circ}<|b|<10^{\circ}$ and  $10^{\circ}<|b|<15^{\circ}$. The results are shown in 
Fig.\ref{fig:index10} and  Fig.\ref{fig:index15}.  One can see that  for large latitudes 
the power-law photon index is nearly  the same  in all directions.  Thus the effect of hardening  of the \gray  spectrum 
is relevant   only to  the Galactic plane.

\begin{table}
\caption{$X_{\rm CO}$ derived from the likelihood fit. } \label{tab:xco} 
\begin{tabular}{lll}
\hline
Ring  &\vline  $X_{\rm CO}$ \\
\hline
$1~\rm kpc<r< 2~\rm kpc$&\vline $1.5 \pm 0.2\times10^{20}$  \\
\hline
$2~\rm kpc<r< 4~\rm kpc$ &\vline  $1.4\pm 0.2\times10^{20}$\\
\hline
$4~\rm kpc<r< 6~\rm kpc$ &\vline  $0.5\pm 0.2\times10^{20}$\\
\hline
$6~\rm kpc<r< 8~\rm kpc$&\vline  $1.4\pm 0.2\times10^{20}$\\
\hline
$8~\rm kpc<r< 12~\rm kpc$&\vline  $2.0\pm 0.2\times10^{20}$\\
\hline
$12~\rm kpc<r< 25~\rm kpc$&\vline  $3.7\pm 0.4\times10^{20}$\\
\hline

\end{tabular}
\end{table}

\subsection {\gray emissivity in the gas rings}
\begin{figure*}
\centering
\subfloat[1 - 2 kpc]{\includegraphics[width=0.3\linewidth]{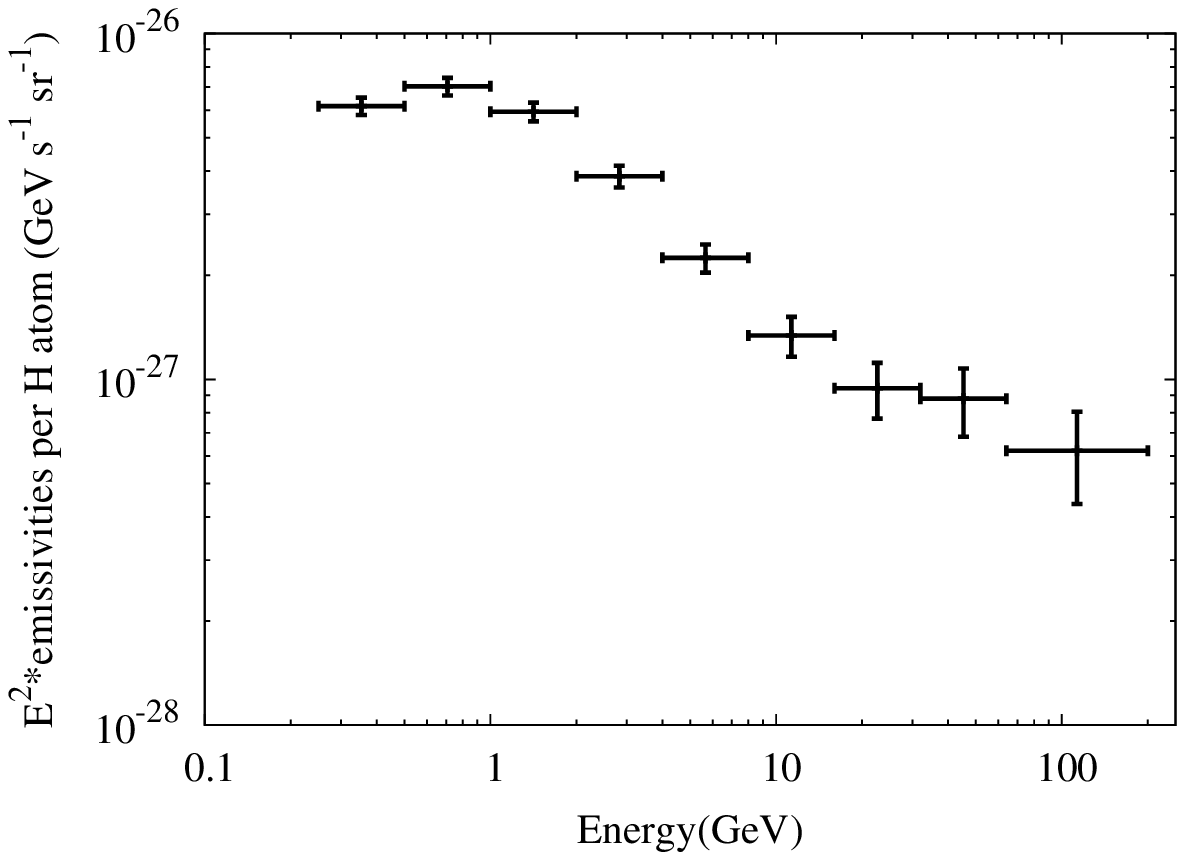}}
\subfloat[2 - 4 kpc]{\includegraphics[width=0.3\linewidth]{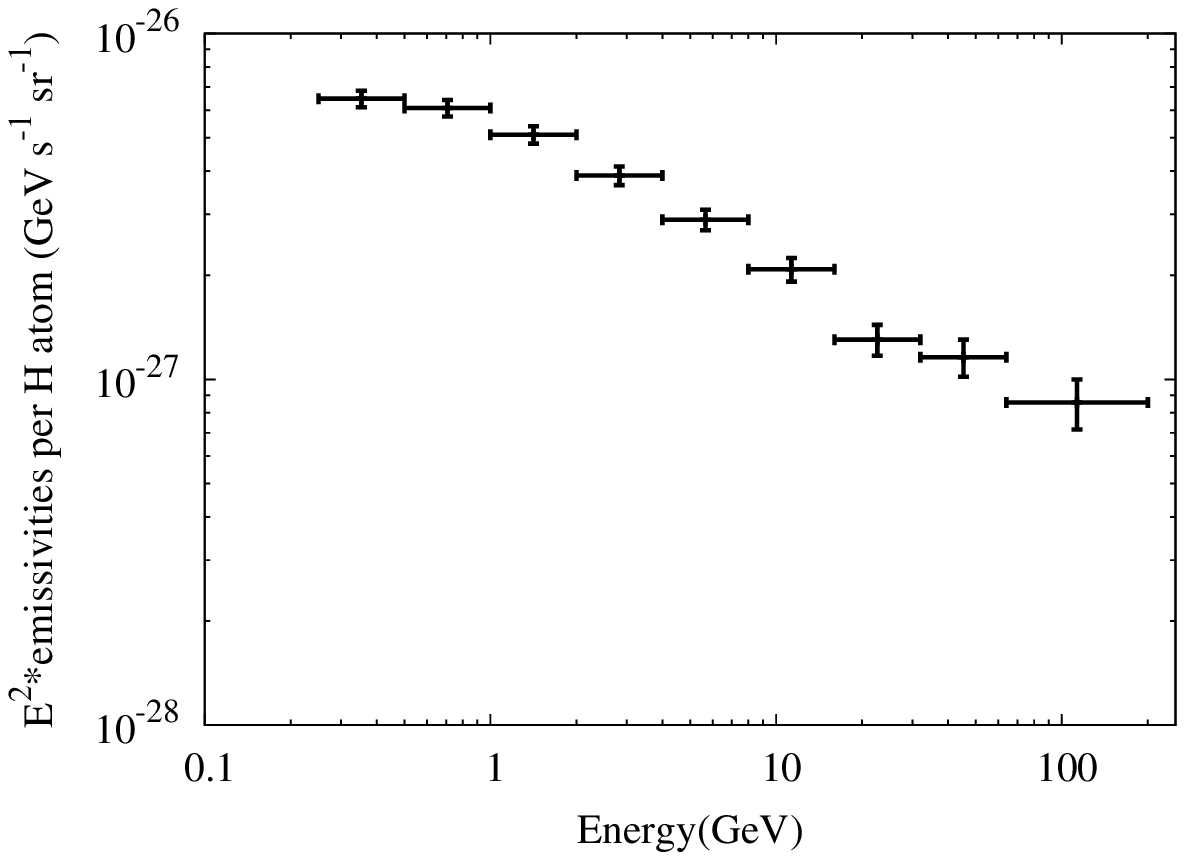}}
\subfloat[4 - 6 kpc]{\includegraphics[width=0.3\linewidth]{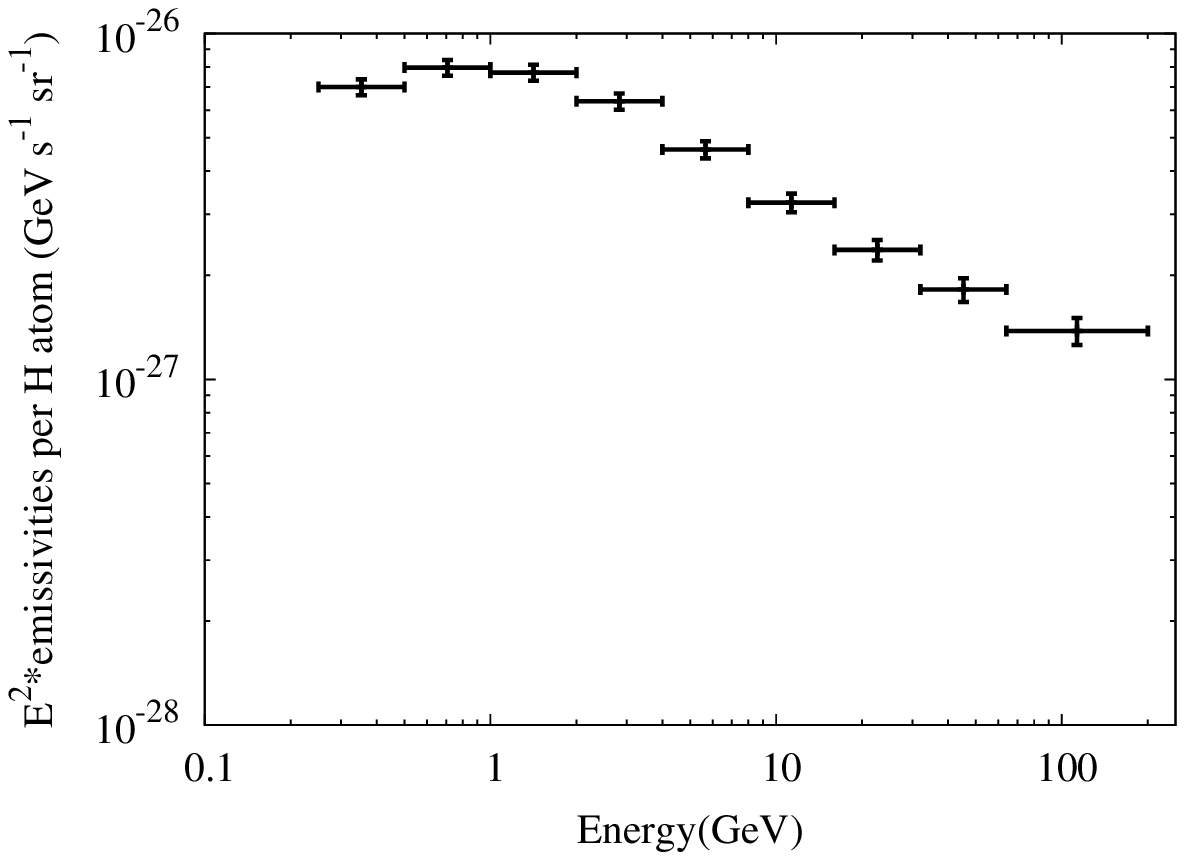}}\\
\subfloat[6 - 8 kpc]{\includegraphics[width=0.3\linewidth]{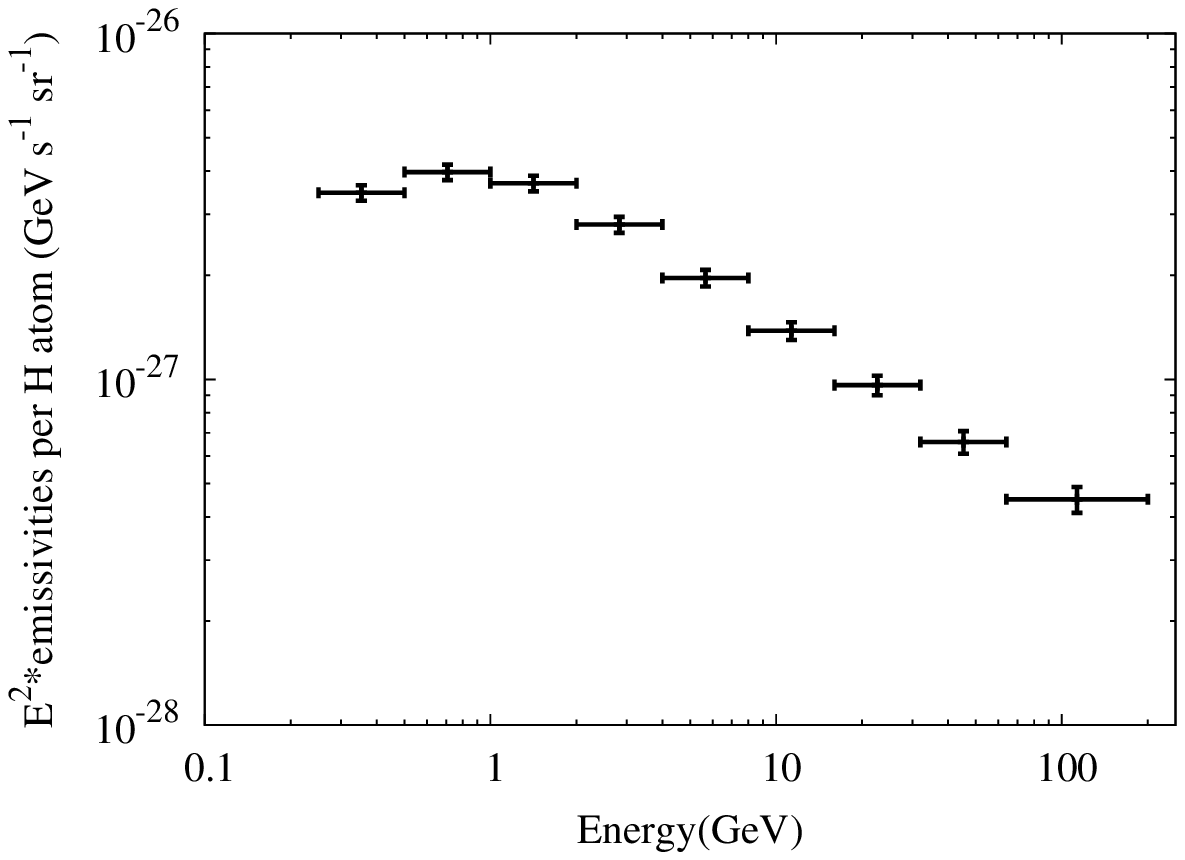}}
\subfloat[8- 12 kpc]{\includegraphics[width=0.3\linewidth]{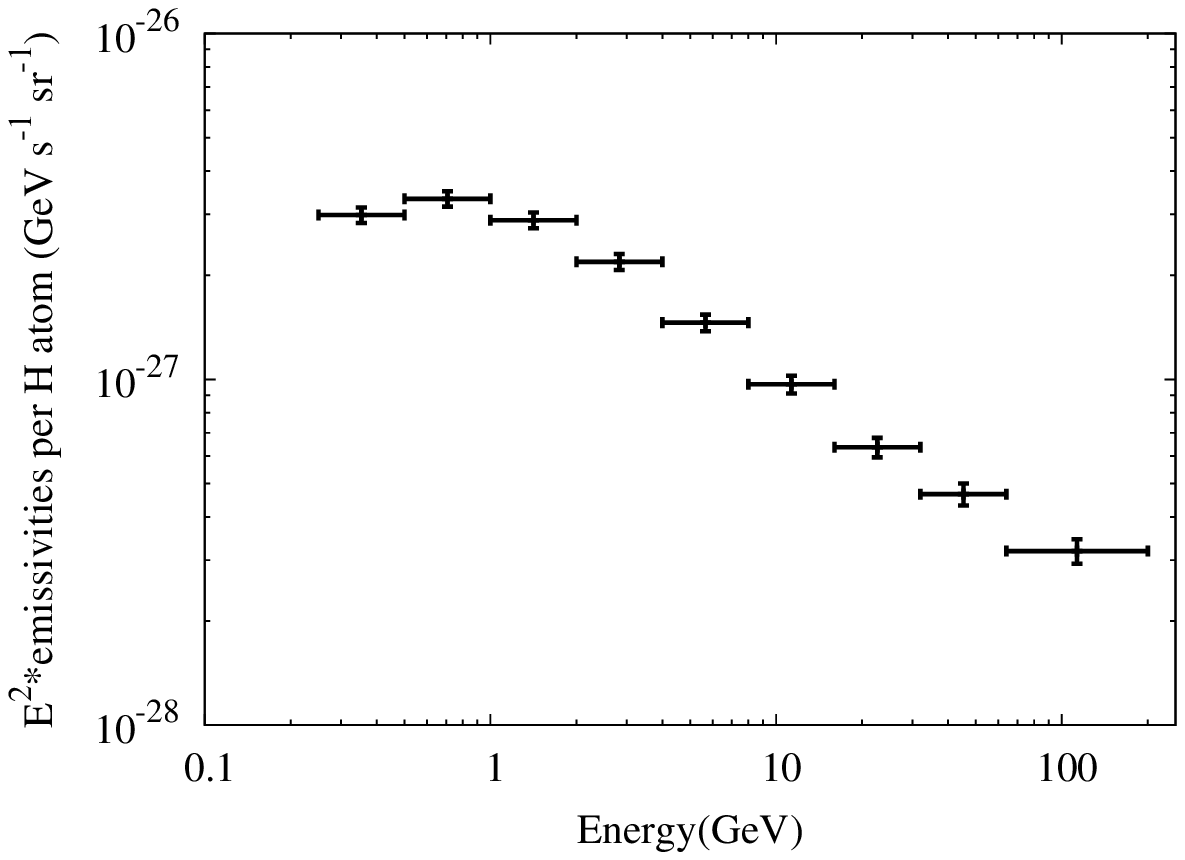}}
\subfloat[12 - 25 kpc]{\includegraphics[width=0.3\linewidth]{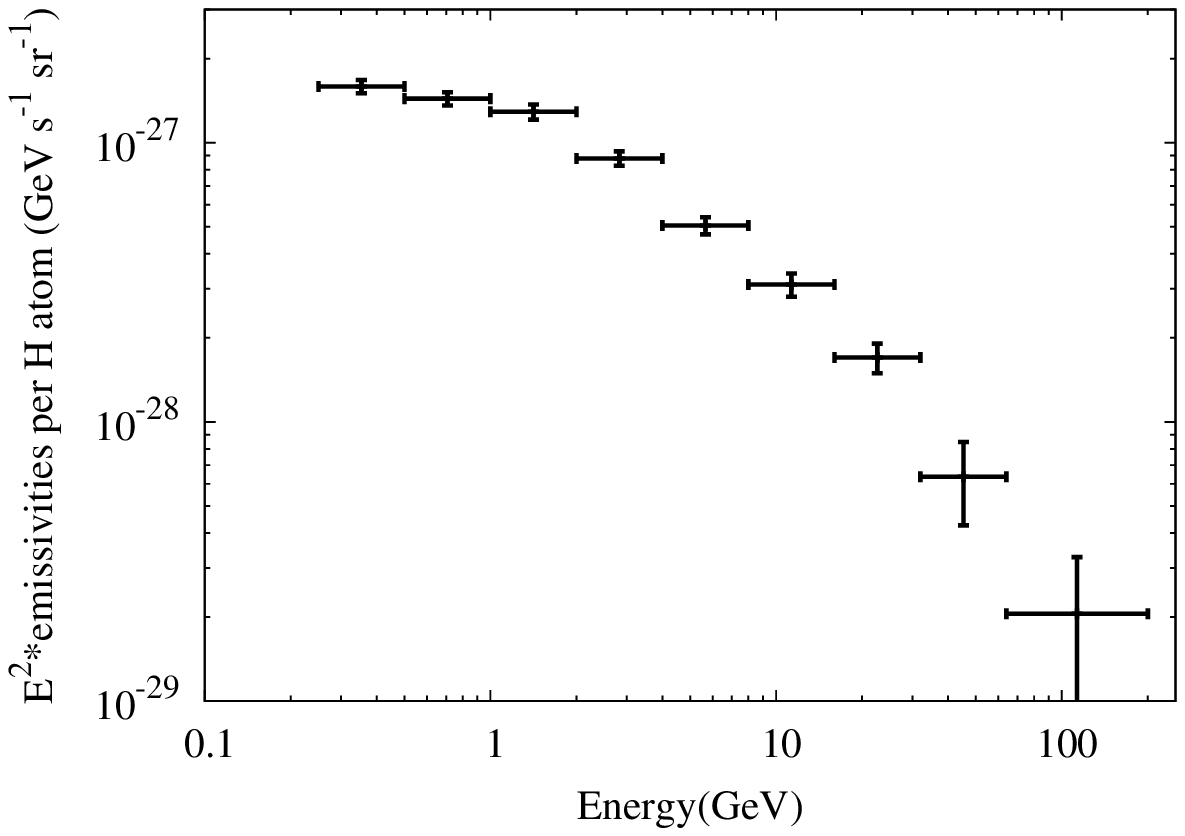}}\\
\centering
\caption{The SED of galactic diffuse \gray emission associated with the gas in different rings around the GC. }
\label{fig:gassed}

\end{figure*}

\begin{figure*}
\includegraphics[width=0.5\linewidth]{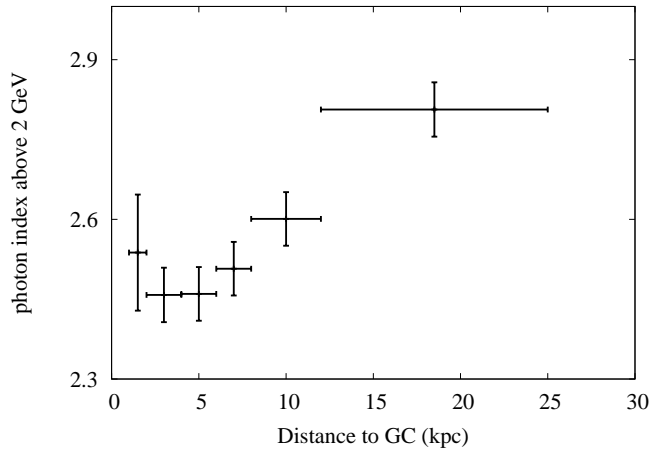}
\caption{The distribution of  the photon index of the galactic diffuse gamma ray emission associated with the 
gas in different rings.}
\label{fig:ringindex}
\end{figure*}

\begin{figure*}
\includegraphics[width=0.5\linewidth]{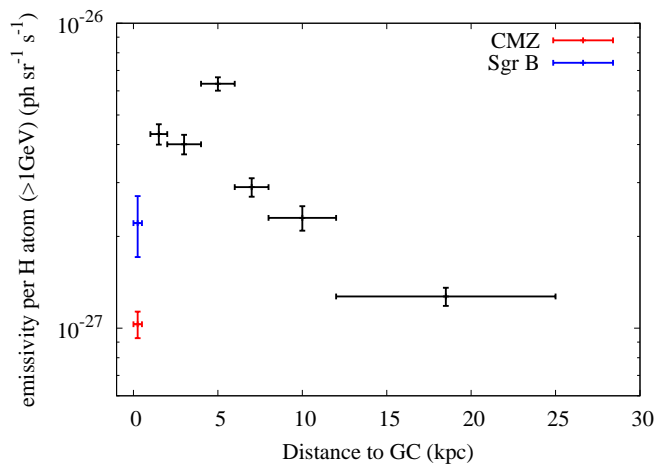}
\caption{Normalised per H-atom gamma ray emissivities above 1 GeV in different rings by using the best fit  conversion 
factor $X_{\rm CO}$   as shown in Tab.1. The points close to the GC are  taken from analysis of the $\gamma$-ray luminosity 
of the entire Central Molecular Zone \citep{fermi_diffuse} and  from the analysis of the emissivity of the giant molecular complex Sgr B2 \citep{yang15}.}
\label{fig:gasemis}
\end{figure*}

To investigated further the tendency of spectral hardening,  we used, instead of dust opacity maps,   the 
HI and CO data  which contain information  on distances. The  data are divided into 6 rings around  the 
GC:  [1,2], [2,4],[4,6],[6,8],[8,12] and [12,25] kpc. To account for the so-called ``dark gas",
 we also included the ``dark'' gas template as described in Sec.2.   To keep the calculation time 
 at  a reasonable level,  we fix  both the absolute fluxes and photon indices of all  point sources to the values 
 from the fit in the dust template fitting, or to the catalog values. The difference in this two cases is small 
 (we included it in the systematic errors of the final data points). To reduce further  the degree of freedom in the likelihood fit,  we first fit the gamma ray emissivity in the 2-4 GeV energy interval, where we deal  with  adequate  photon statistics combined with a reasonably good  PSF. From the results of fitting in this  energy band,  we derive the factor $X_{\rm CO}$  by assuming that the  \gray emissivity per $H_2$ is twice of the emissivity  per neutral hydrogen atom . Then we fix this factor and apply the likelihood fitting for other energy intervals.  The results are shown in Fig.\ref{fig:gassed}, Fig.\ref{fig:ringindex} and Fig.\ref{fig:gasemis}. The fitted $X_{\rm CO}$ are listed in Table.\ref{tab:xco}. 

\begin{figure*}
\includegraphics[width=0.5\linewidth]{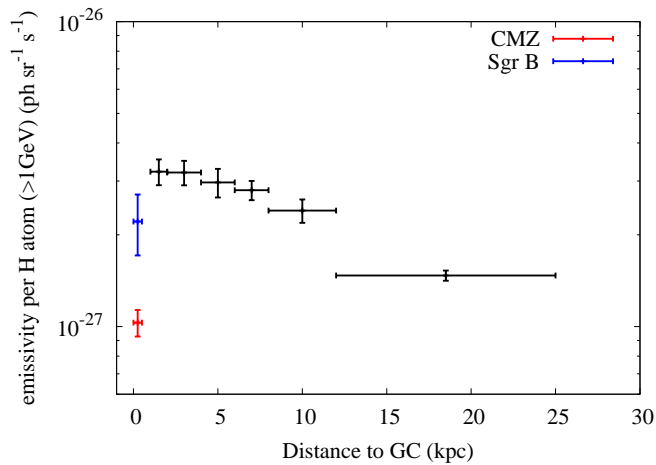}
\caption{The same as in Fig.\ref{fig:gasemis} but for  the conversion factor
 $X_{\rm CO}$  fixed at  the local value $2.0\pm 0.2\times10^{20}$ .}
\label{fig:gasemisfix}
\end{figure*}

The results are consistent with the conclusion based on the analysis  of dust templates described in the previous section. The derived power-law indices imply  a clear  tendency of spectral  softening toward the outer Galaxy. Namely,  the power-law photon index above 2~GeV  varies from $\approx$2.4 in the innermost ring to 2.8 in the outermost one. Furthermore,  it can be seen in Fig.\ref{fig:gasemis} that  the  normalised  \gray emissivity (per H-atom) also varies with the distance to the GC. Since the normalised \gray emissivity is  proportional to the CR density,  Fig.\ref{fig:gasemis} can be treated as a radial profile of  the CR density.  One can see that the CR density achieves its maximum in the ring [4,6] kpc.  Note that  the peak is coincident with the peaks of  distributions of SNRs \citep{case98} and OB stars   \citep{bronfman00}.  Interestingly,  
the peak coincides  with  the  minimum of the  conversion factor $X_{\rm CO}$  which appears in the same ring  
(see Table.\ref{tab:xco}).  Note that the small values of the  $X_{\rm CO}$ factor in the inner Galaxy
has been independently derived  in   \citet{fermi_diffuse_old}.  
The observed gradient in metallicity across the Galaxy \citep{cheng12}  also supports  the    
low  values  of $X_{\rm CO}$. Yet,  such a coincidence between  the CR density and  the conversion factor 
$X_{\rm CO}$ seems, to some extent,  suspicious.  Thus,  one cannot exclude that the peak in the \gray emissivity  could be caused by  the underestimation of the  gas density in this region.   For  comparison,  in fig.\ref{fig:gasemisfix}  we 
show  the  \gray emissivity derived under an {\it ad hoc}  assumption of  constant  conversion factor $X_{\rm CO}$ 
in all rings fixed to  the  value $2.0\pm 0.2\times10^{20}$.  One  can see that in this case the maximum in the 
$\gamma$-ray emissivity in the 4-6 kpc ring disappears.

\begin{figure*}
\includegraphics[width=0.5\linewidth]{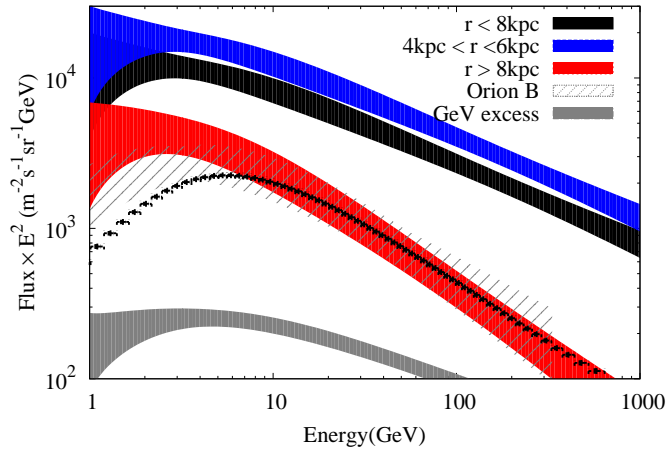}
\caption{The CR proton spectra  in the inner  ($r \leq 8$~kpc) and outer ($r \geq 8$~kpc) regions, as well as 
in the 4-6 kpc ring  derived from 
$\gamma$-ray  emissivities  presented in Fig.\ref{fig:gasemis}. Also are  shown the proton spectra derived from the 
$\gamma$-ray  measurements of the nearby molecular cloud Orion B \citep{yang14}. 
and from the low-energy  $\gamma$-ray component called ``GeV excess".  The direct measurements of the 
CR proton spectrum are from  the  AMS-02 collaboration report \citep{aguilar15}, which are shown as black squares. 
 }
\label{fig:proton}
\end{figure*}

\begin{table}
\caption{Energy density of CRs  in different rings. } \label{tab:cre} \centering
\begin{tabular}{lll}
\hline
Ring  &\vline  $\epsilon_{\rm CR}$ (1 - 10GeV) ($\rm eV/cm^3$)  &\vline  $\epsilon_{\rm CR}$ (>10GeV) \\
\hline
$r< 8~\rm kpc$&\vline~~ $1.09 \pm 0.30$ &\vline~~$0.67 \pm 0.13$ \\
\hline
$r> 8~\rm kpc$ &\vline ~~  $0.37 \pm 0.14$ &\vline~~$0.12 \pm 0.04$ \\
\hline
$4~\rm kpc<r< 8~\rm kpc$ &\vline~~  $1.60 \pm 0.45$ &\vline~~$1.00 \pm 0.15$ \\
\hline
AMS-02 &\vline~~~$0.13 \pm 0.001$ &\vline~~$0.09 \pm 0.001$ \\
\hline
GeV excess &\vline  ~~ $0.02 \pm 0.01$ &\vline~~$0.02 \pm 0.01$ \\
\hline

\end{tabular}
\end{table}

\section{Discussion}
In this paper we present the results  of our analysis of the diffuse galactic $\gamma$-ray emission  
based on the  \fermi  \gray  data  combined with  the HI, CO and the dust-opacity maps.  The  energy range of 
detected $\gamma$-rays  from  100 MeV to 200~GeV  allows 
derivation  of  detailed spectral and spatial  distributions of CRs  over almost
four energy decades, from   mildly relativistic (sub-GeV) to ultra-relativistic  (multi-TeV)  energies. 

The results   described in the previous sections demonstrate  strong variations of  both the energy  
spectra and  the absolute fluxes of $\gamma$-rays throughout the  entire galactic plane  
($0^{\circ}<l<360^{\circ}$ and $|b|<5^{\circ}$). The energy spectra 
of $\gamma$-rays arriving from the directions of inner  Galaxy appear significantly 
harder compared to the spectra of radiation from outer parts of the Galaxy.  
The  tendency  of  the spectral change  is clearly seen in Fig.\ref{fig:dustsed} and  Fig.\ref{fig:dustindex}.  
This conclusion  agrees with the recent analysis of the Fermi LAT data  reported in  ref. \citet{fermi_diffuse}, 
and, with some reservation,  with the old results reported by the  EGRET 
team \citep{egretplane} \footnote {Note  that although 
the EGRET result later has been criticised, and the very 
existence of the ``GeV bump"  has been  discarded by the community, one should point out  that in general terms 
the latter is in reasonable agreement with the recent  \fermi  data.}.  A  hard diffuse $\gamma$-ray spectrum   
has been  recently reported  also  in \citet{neronov15} who claimed a 
universal photon index  close to 2.45  throughout  the entire  Galactic Disk. However,  this conclusion
based on a rather limited range of galactic longitudes ($|l| \leq 90^\circ$), is misleading.    The results shown in 
Fig.\ref{fig:dustsed} and Fig.\ref{fig:dustindex}   reveal  a non-negligible spectral variation throughout the 
galactic plane.  In particular, while   the  $\gamma$-ray  spectrum   in the  inner  parts of the Galaxy 
is significantly harder than the spectrum of locally measured  CRs, the latter  is quite close 
to the  spectrum  of $\gamma$-rays (and, consequently,  to the spectrum of parent protons) 
from  the anti-centre  direction (where the Solar system is located).  
Moreover,  the comparison of Fig.\ref{fig:dustindex} with Fig.\ref{fig:index10} and Fig.\ref{fig:index15}, 
makes it clear that the spectral hardening  takes place  only in the galactic plane.  

In general, the spectral hardening of  $\gamma$-rays could be caused  by contamination of the truly diffuse  flux of 
$\gamma$-ray (i.e. the ones  produced in  interactions of CRs)  by  discrete  $\gamma$-ray 
sources concentrated in the inner Galaxy.  We believe that the  discrete sources  have been 
carefully  treated in our analysis. Nevertheless,   we cannot exclude  the possibility that  a new 
population of weak but numerous  hard-spectrum \gray sources, 
which have not been resolved by  \fermi,   significantly contribute  to the truly diffuse $\gamma$-ray background.   

Alternatively, the hardening could be caused, in principle,  by undervaluation  of  the contribution of the 
IC component of $\gamma$-rays in the inner Galaxy. In this work,  a spatial IC template  based on the 
calculations with the  GALPROP code \citep{galprop},  was  used  to model the IC emission.  
Although  the interstellar radiation fields (ISRF) are  poorly constrained,
their enhancement in the inner Galaxy (compared to the values provided by  GALPROP)  
hardly can exceed a factor of two or three. Meanwhile,  at multi-GeV energies  
the component of $\pi^0$-decay $\gamma$-rays strongly, by on order of magnitude, dominates  over the 
IC contribution calculated for the ``standard"  fluxes of ISRF \citep[see][]{aa00}.  Consequently, the interpretation  of the spectral hardening of $\gamma$-rays by  
an enhanced  IC contribution, seems rather unlikely.   

Thus,  we may conclude, although  not without  the  caveat  concerning  the   
possible non-negligible contribution of unresolved $\gamma$-ray sources, 
that the major fraction of the  diffuse $\gamma$-ray flux   at GeV energies 
arises  from  interactions of  CR protons and nuclei with the interstellar gas (through the production 
and decay of  the secondary neutral pions). In the  framework of this interpretation, the hardening 
of the $\gamma$-ray spectrum in the inner Galaxy can be explained by  the concentration of  CR accelerators in the 
inner Galaxy  and/or  by the CR propagation effects \citep[see e.g.][]{gaggero15}.  The radial distributions of  the normalised 
$\gamma$-ray  emissivity  (Fig.\ref{fig:gasemis} and Fig.\ref{fig:gasemisfix}), as well as the radial dependence  
of  the  photon index  (Fig.\ref{fig:ringindex}),  
contain  unique information regarding the spatial distribution of CR accelerators.   

If the distinct  maximum  in the radial distribution of the $\gamma$-ray emissivity  in Fig.\ref{fig:gasemis}  
calculated for the values of the conversion factor  $X_{\rm CO}$  from  Table 1,  is  real, 
it is straightforward to assume 
that the CR accelerators  are concentrated  in the 4-6 kpc ring.  The continuous injection of CRs into the interstellar 
medium  can explain the  gradual drop of the $\gamma$-ray emissivity in both directions - towards the GC and to the 
periphery  of the Galaxy (see Fig.\ref{fig:gasemis}).  While in the  homogeneous and spherically symmetric medium we expect $1/r$ 
type distribution  for  the CR density \citep[see e.g.][]{aa00},   radial dependence of the diffusion coefficient 
and  its  possible anisotropic character  could result in a  deviation 
from the $1/r$  dependence.   Detailed numerical  treatment of the CR propagation,  and 
the comparison  of  theoretical predictions with  the results presented in Fig.\ref{fig:gasemis} and  Fig.\ref{fig:ringindex}, 
would provide an important information  on the  spatial distribution of sources of CRs,  on the character of their propagation in the interstellar magnetic fields,  on the total CR injection  rate,  {\it etc}.

What concerns the harder  energy spectrum 
of  CRs  which are currently confined in the 4-6 kpc ring,   compared to their
spectrum in outskirts of the Galaxy,  a  possible  reason of this effect 
could be  an intrinsic feature of particle accelerators in the ring. For example, one may speculate that  
the sources in the inner Galaxy  accelerate  CRs with harder spectra  than  the CR accelerators do in the outer Galaxy. 
An alternative  explanation of of this effect could be related to the specifics of propagation of particles  in the inner Galaxy. In particular, it  could be caused by different rates of escape of CRs  in different parts of the Galaxy.
For example, the stronger galactic wind in the inner Galaxy, caused by  a  higher local pressure, 
may result  in the transport of  CRs  which at low energies  might  be dominated by advection rather than diffusion. Consequently, at low energies  the  shape of the CR  spectrum in the inner Galaxy would not suffer 
deformation \cite[]{aa00}, while in the outer Galaxy the energy-dependent diffusion could lead to significant 
steepening of the original (acceleration) spectrum.  Apparently, these two  scenarios  need 
further observational and theoretical  inspections. In particular, it would be important to perform independent measurements of  the CR spectra from individual massive clouds  which can serve as CR  barometers \citep{aharonian91,aharonian01,casanova10}. Such measurements, already have been  conducted for the nearby molecular clouds in the Gould Belt \citep{yang14} and local H{\sc i} regions \citep{fermi_h1, casandjian_lis}. They  have  
revealed  absolute fluxes and  energy spectra of CRs which are in a good 
agreement with  the results of this paper for the outer Galaxy,  and with the recent direct measurements of CR protons 
reported by the AMS-02 collaboration \citep{aguilar15} (see Fig.\ref{fig:proton}). 
 
Finally,  we note that  the 4-6 kpc ring seems to be a rather natural place for 
concentration of CR accelerators, given that the 
potential  CR source populations, such as pulsars, OB stars and SNRs,  have radial profiles which also 
shows a maximum  at several kpc  from the Galatic Centre \citep{case98,bronfman00}.   
Yet,   it sounds  somewhat  suspicious that the reason of 
the maximum in the $\gamma$-ray luminosity in the 4-6 kpc ring is  the  radial dependence of the 
conversion factor $X_{\rm CO}$  which  has a sharp minimum exactly in the same 4-6 kpc  ring (see Table.\ref{tab:xco}).   
Whether we deal with  an accidental coincidence or there are some  deeper reasons for such a coincidence, 
this is a question  which is beyond the  framework  of the present  paper.    On the other hand, if we assume a smoother 
radial-dependence  of the  conversion  factor $X_{\rm CO}$, the corresponding radial distribution of 
the $\gamma$-ray emissivity would become  quite different from the distribution in Fig.\ref{fig:gasemis}.
For example,   the  ({\it ad hoc} assumed)  constant  conversion factor $X_{\rm CO}$ 
would remove the maximum  of the  $\gamma$-ray emissivity  at 4-6 kpc (see Fig.\ref{fig:gasemisfix}). 
Instead,  it would result in 
a continuous increase of the CR density towards the GC.   In this case, the GC itself  would  
be the most natural  place for CR production, given that  some other  \gray phenomena like the 
TeV $\gamma$-ray emission of the Central Molecular Zone 
\citep{hessgc06, hessgc16},  
the  so-called ``GeV excess"  \citep{goodenough09, ak12, daylan13, macias14, weniger14}, the   
Fermi Bubbles \citep{su10},  are most likely linked, in one way or another, to the GC. 
To sustain the observed density of CRs,  the average CR production over the last $10^7$ years should 
proceed at the rate of $10^{41}$erg/s. The only source  in this  compact region 
which could  provide (in principle) such a power in relativistic particles 
is  the   Supermassive Black Hole in the dynamical centre of our Galaxy.
In this scenario, the significant reduction of the CR density in the very central (within 100 pc) region of the GC 
can be explained by  the  rather low activity of the source over the last $10^{3-4}$ years (the characteristic 
timescales of  propagation of CRs  on100~pc  scales), and/or  the fast removal of CRs from this region, 
for example,  due to convection.   

Independent of the origin of the parent protons,  the results of this paper demonstrate that the concept of the so-called ``sea"  of galactic CRs,
which assumes nearly uniform spatial and energy distributions of  CRs  in the Galactic Disk,  is a very 
rough approximation  which  perhaps can be used for  zeroth-order estimates of  global  parameters, 
but not for a  quantitative description of the  CR production and transport throughout the Galaxy.
This is demonstrated in Fig.\ref{fig:proton} where we show the average energy spectra of CR protons at different radial 
distances from the GC. They are derived  from  $\gamma$-ray emissivities  presented  in Fig.\ref{fig:gasemis} (in calculations 
we  take into account  the enhancement of the $\gamma$-ray production rate due to the CR nuclei  
by  a  factor of 1.8 \citep{kafexhiu14}). 
In the same figure we show the CR proton fluxes  in our neighbourhood  
from direct measurements of AMS-02 \citep{aguilar15}, and from the  $\gamma$-ray  observations in the nearby 
molecular cloud Orion B2 \citep{yang14}.  The flux of the CR proton component 
responsible for the so-called ``GeV-excess" in the central several kpc region of the inner Galaxy is also shown. 
Note that at  low  energies  we may expect  significant contribution from the bremsstrahlung and IC components 
of $\gamma$-rays,  therefore the  curve  should be considered as an upper limit. In any case, the contribution of  
particles responsible for the ``GeV-excess'' is well below  the contribution of the main CR component.

\bibliography{ms}
\end{document}